\documentstyle[epsfig,11pt]{article}
\def \blankline{\vspace{0.4 cm}}

\begin{document}
\begin{center}
{\large\bf Detection of Gamma-Ray Bursts in the 1 GeV-1 TeV energy 
range by ground based experiments}

\blankline
\blankline

Silvia Vernetto

{\it\small 
Istituto di Cosmogeofisica del C.N.R., Corso Fiume 4, 10133 Torino, Italy} 
\end{center}

\blankline
\blankline
\blankline
\blankline

{\bf\large Abstract\\}

\blankline

Ground-based extensive air shower arrays can observe
Gamma Ray Bursts in the 1 GeV - 1 TeV energy range 
using the ``single particle" technique.
The sensitivity to detect a GRB as a function of the burst
parameters and the detector characteristics are discussed.
The rate of possible observations is evaluated,
making reasonable assumptions on the high energy emission, the
absorption of gamma-rays in the intergalactic space, the
distribution of the sources in the universe 
and the bursts luminosity function.
We show that a large area detector located at high mountain altitude 
has good prospects for positive detections, providing
useful information on the high energy component of GRBs.

\blankline
\blankline
\blankline
\blankline
\blankline

e-mail: vernetto@lngs.infn.it

Tel.: +39-011-6306810; Fax: +39-011-6604056.

\newpage

\section {Introduction}

Thirty years after their discovery Gamma-Ray Bursts (GRBs) 
still remain one of the most intriguing mystery of the universe.
The succesful observations of BATSE aboard the Compton Gamma Ray Observatory
and the exciting detections in X, optical and radio bands during the
last two years put an end to the discussions about their distance,
proving that the sources of GRBs (or at least most of them)
are cosmological objects.
Supposing an isotropic emission, the measured distances
imply an amount of radiated energy 
as large as $\sim 10^{51} - 10^{54}$ erg, setting GRBs among the most 
powerful astrophysical phenomena in the universe.
According to most theoretical models, gamma-rays can be produced
by the synchrotron radiation of charged particles accelerated in the
shock wave of a fireball, the relativistic explosion due to
a still unknown catastrophic event, presumably the formation of a black
hole through a coalescence of a compact binary system or a 
``hypernova" \cite{1}.

GRBs have been well studied in the KeV-MeV energy range, where they show a 
spectrum with a power law tail, in agreement with synchrotron models.
The extension of measurements in the GeV-TeV range would be
of major importance, since it could impose strong constraints on 
the physical conditions of the emitting region.
Theoretical predictions of the high energy emission in 
relativistic fireball models are given in \cite{2,3,4}.
Unfortunately the gamma-rays absorption in the intergalactic space 
by interaction with starlight photons 
prevents the study of GRBs above $\sim$ 1 TeV.

So far, only 7 GRBs have been observed at energy $E >$ 30 MeV
by EGRET on the Compton Gamma Ray Observatory,
and in 3 of them photons of energy $E >$ 1 GeV have been detected \cite{5}.
The observed spectra are consistent with a power law behaviour
with no high energy cutoff up to the maximum energy observable
by the instrument.
It's important to note that these 7 GRBs are as well the most 
intense events observed by BATSE at energy E $>$ 300 KeV;
taking into account the small field of view of the
EGRET detector ($\Omega \sim$0.6 sr) and its limited sensitivity,
the small number of high energy detections does not contradict
the idea that all GRBs could have a high energy component.

Less constrained in size than space born detectors,
ground based experiments can explore the energy region
above $\sim$10 GeV,
measuring the secondary particles generated by gamma-rays interactions
in the atmosphere ($e^{\pm}$ or Cerenkov photons) that reach the ground. 
Cerenkov telescopes detect the Cerenkov light emitted by $e^{\pm}$
in the high atmosphere. Although several successful observations 
of steady gamma-ray sources at energies of a few hundreds GeVs
have been performed with Cerenkov measurements,
the small fields of view of telescopes (few squared degrees) and 
the limited live time (constrained by weather and darkness conditions to
be of the order of 10$\%$)
make the Cerenkov technique not suitable for observations of transient and
unforeseeable events as GRBs (the probability
for a GRB to happen by chance in the field of view of a telescope is less than 
10$^{-4}$).
However, while observations of the burst onset
are highly improbable, delayed observations are feasible and 
they could be of great interest \cite{6}
(recall the delayed high energy emission observed by EGRET
more than one hour after the famous event GRB940217 \cite{7}).

Air shower arrays detect the charged particles (mostly $e^{\pm}$)
of the showers reaching the ground;
they have a much larger field of view (about $\pi$ sr) and 
a live time of almost 100$\%$.
In the next section we will discuss the possibility to 
detect GeV-TeV photons from GRBs using air shower arrays,
in particular using the ``single particle" technique.
In section 3 the expected rate of positive observations will be evaluated, 
making reasonable assumptions on the
luminosity and distance distribution of GRBs.

\section{The ``single particle" technique}

The simplest way to observe the high energy component of GRBs 
with a ground based experiment is using an air shower array 
working in ``single particle" mode. Air shower arrays, 
made of several counters spread over large areas, usually detect
air showers generated by cosmic rays of energy $E >$ 10-100 TeV;
the arrival direction of the primary particle is measured by 
comparing the arrival time of the shower front 
in different counters, and the
primary energy is evaluated by the number of secondary particles
detected.
Air shower arrays can be used in the 
energy region E~$<$~1~TeV working in single particle mode,
i.e. counting all the particles hitting the individual detectors, independently
on whether they belong to a large shower or they are the lonely
survivors of small showers. Because of the cosmic ray spectrum steepness,
most of the events detected in this mode of operation
are in fact due to solitary particles from small 
showers generated by 10-100 GeV cosmic rays.

Working in single particle mode an air shower array could in principle 
detect a GRB in the energy range 1-10$^3$~GeV, 
if the secondary particles generated by
the gamma-rays give a statistical significant excess of events
over the all-sky background due to cosmic rays.
The beauty of this technique consists in its extreme simplicity:
it is sufficient to count all the particles hitting the detectors
during fixed time intervals (i.e. every
second, or more frequently depending on the desired time resolution)
and to study the counting rate behaviour versus time,
searching for significant increases; the observation of an excess
in coincidence with a GRB seen by satellites would be an
unambiguous signature of the nature of the signal.

The single particle technique does not allow the 
measurement of the energy and the direction of the primary gamma-rays, 
because the number of particles (often only one per shower) is too small 
to reconstruct the shower parameters. However it can allow
the study of the temporal behaviour of the high energy emission, 
and, with some assumptions on the
spectral slope (possibly supported by satellite measurement at lower energy)
it can give an evaluation of the total power emitted at high energy.

Fig.1 shows the mean number $n_e$ of electrons and positrons
reaching the ground at different altitudes $h$ above
the sea level generated by a gamma-ray, as a function of the gamma-ray
energy, according to the Greisen analytical expression \cite{8}.
The curves are given for two different zenith angles of the primary
gamma-ray, $\theta$ = 0$^{\circ}$ and $\theta$ = 30$^{\circ}$.
The number of particles
strongly increases with the altitude, in particular at low energies,
and decreases at high zenith angles, in particular at low altitude.
Actually, due to the gamma-rays photoproduction, 
a small number of muons is produced as well; 
however, in the energy range $E=1-10^3$ GeV,
the number of muons is negligible compared with the 
number of electrons and positrons (except for $h<$1~km, where they
are comparable) and they will not be considered in the following
discussion.

A possible way to increase the number of detectable particles is
to exploit the air shower photons, that are much more numerous than $e^{\pm}$.
Fig.2 shows the ratio $r_s=n_{ph}/n_e$
of the mean number of photons and the mean number of $e^{\pm}$,
as a function of the altitude.
The values have been obtained simulating electromagnetic cascades 
induced by gamma-rays with a zenith angle $\theta$ = 30$^{\circ}$.
The detector is assumed with an energy threshold $E_{th}$=3 MeV.
The ratio $r_s$ range from $\sim$6 to $\sim$11 depending on the altitude
and on the primary gamma-ray energy. The conversion
of a fraction of photons in charged particles (by pair production
or Compton effect) would increase the detectable signal. 
Placing a suitable layer of lead over the detector, a small fraction of 
$e^{\pm}$ would be absorbed,
but a large number of photons would interact and 
increase the GRB signal, by a factor depending on the converter
geometrical characteristics.
An interesting possibility to
avoid the absorption of $e^{\pm}$ is to adopt a thick scintillator
detector: it would work both as a target for photons and as a detector
for charged particles \cite{9}.

\subsection{The background}

The background is due to secondary charged particles
from cosmic rays air showers. The rate depends on the altitude $h$ 
and, to a lesser extent, on the geomagnetic latitude $\lambda$ of the observer. 
The latitude effect
is almost negligible at sea level and increases with the altitude:
at $h$ = 5 km a.s.l. the
background at $\lambda$ $>$ 45$^{\circ}$ is about 1.6 times larger
then near the magnetic equator \cite{10}.

The dependence of the background on the altitude is more important.
Fig.3 shows the flux of $e^{\pm}$, $\mu^{\pm}$ and charged hadrons
as a function of the altitude above the sea level,
for an observer located near the geomagnetic equator,
obtained simulating with the CORSIKA code
the atmospheric cascades generated by protons and Helium nuclei
with an energy spectrum according to \cite{11} 
and using a dipole model for the geomagnetic field.
The detector energy threshold is $E_{th}$=3 MeV.
The total background rate ranges from $\sim$250 to 2000 events m$^{-2}$ s$^{-1}$
increasing the altitude from 0 to 5000 m.
The rates obtained by the simulations are in good agreement with the
counting rates measured at 2000 m \cite{12} and 4300 m \cite{13}.
Most of the background is due to muons for 0 $< h <$ 3.5 km and to
$e^{\pm}$ at higher altitude
(note that a detector able to separate the electron and the muon 
components would allow the reduction of the background by 
the rejection of muons).

It's worth noting that the number of $e^{\pm}$ generated by a primary 
gamma-ray (see Fig.1) increases much more rapidly with the altitude than the 
background does.
This is due to two factors: {\it 1)} given a primary energy, the
electromagnetic shower generated by a proton penetrates more deeply in the 
atmosphere than the gamma-ray shower does, {\it 2)} 
the $\mu^{\pm}$ flux, that strongly 
contributes to the background rate, is less dependent on the altitude 
than the $e^{\pm}$ flux.

The possible conversion of secondary photons in detectable
particles (introduced in the previous section as a method
to increase the GRB signal) would produce an increase in the background as well.
Fig.3 also shows the number of photons reaching the ground 
due to cosmic rays. 
However the ratio $r_b$ of the number of photons and the
number of charged particles  is 
considerably smaller than the corresponding value $r_s$ for primary 
gamma-rays ($r_b <3$ at any altitude, while $r_s \sim$ 6-11, as shown in Fig.2).

The background rate is not strictly constant and several mechanisms
are responsible for variations on different time scales
with amplitudes up to a few percent.
First of all, variations in the atmospheric pressure
affect the secondary particle flux:
an increase (decrease) of the ground level atmospheric pressure
results in a reduction (enhancement) of the background rate,
because of the larger (smaller) absorption of the electromagnetic
component. The pressure and the background rate are linearly
anti-correlated (e.g. the correlation coefficient
between the counting rate percent increase and the pressure variation
measured at 2000 m a.s.l. is $\sim -0.5\%$ per millibar \cite{14}).
By monitoring the pressure at the detector position it is possible
to correct the data for this effect.

The 24 hours anisotropy (due to the rotation of the Earth in the
interplanetary magnetic field), and the solar activity (e.g. large solar flares)
modulate the low energy primary cosmic ray flux.
Although the amplitudes of these variations are not negligible
(up to a few percent of the counting rate in some cases)
the time scales of these phenomena (hours) are too long to simulate
short duration events as GRBs.

Shorter variations in the single particle counting rate have been measured
in coincidence with strong thunderstorms and have been ascribed to the
effects of atmospheric electric fields on the secondary particles flux
\cite{14,15}; however the occurrences of these events are very rare
and also in this case the observed time scales
($\sim$ 10-15 minutes) are longer than the typical GRB duration.

From the experimental point of view,
more troublesome are possible instrumental effects, such as electrical noises,
that could simulate a GRB signal producing spurious increases 
in the background rate.
Working in single particle mode require very
stable detectors, and a very careful and continuous monitoring of the
experimental conditions.
By comparing the counting rate of the single detectors 
(i.e. the scintillator counters, in the case of an air shower array)
and requiring simultaneous and consistent variations of the rate
in all of them, it is possible
to identify and reject the noise events due to instrumental effects.

\section{Sensitivity to detect high energy GRBs}

The aim of this section is to quantify the
actual possibilities of observing 
high energy GRBs using the single particle technique.
First, we present a simple model of high energy emission, 
in order to define a GRB using a limitate set of parameters.
Subsequently, the response of the detector is evaluated
as a function of the burst parameters and detector features.
We conclude with an estimation of the possible rate of positive observations.

\subsection{Parametrization of the high energy emission}

In this simple model a GRB is assumed to release a 
total amount of energy $L$
 in photons of energy E $>$ 1 GeV;
the emission spectrum is assumed constant during the 
emission time $\Delta t_e$ and represented by 
$dN/dE \ = \ K \ E^{-\alpha }$ (number of photons emitted 
per unit energy) up to a cutoff energy $E_{max}$.
The value of $E_{max}$ depends on the emission processes and the
physical conditions at the source; 
a sharp cutoff at $E_{max}$ is probably unrealistic, 
but for our purposes such a simple model is sufficient.
The total energy $L$ and the spectrum are related by
$L=K\int_{1~GeV}^{E_{max}}E^{-\alpha+1}dE$.

If the GRB source is located at a cosmological distance corresponding
to a redshift $z$, the observed burst duration is
$\Delta t=\Delta t_e (z+1)$ due to time dilation,
and the shape of the observed spectrum is determined by two factors:

\noindent
$a)$ the photons energies
are redshifted by a factor ($z+1$) due to the expansion of the universe
(the spectrum slope $\alpha$ doesn't change);

\noindent
$b)$ the high energy gamma-rays are absorbed in the intergalactic space through
$\gamma + \gamma \rightarrow e^+ e^-$ pair production
with starligth photons; 
the probability of reaching the Earth for a photon emitted by a source
with a redshift $z$ is $e^{-\tau(E,z)}$, where 
$\tau$ increases with $E$ and $z$.
The absorption becomes important
when $\tau >1$; according to \cite{16}, $\tau$ reaches the
value of unity at $z\sim$0.5 for $E$=100 GeV, and at $z\sim$2 for $E$=20 GeV.
Fig. 4 shows a spectrum with a slope
$\alpha=2$ affected by the absorption, for different $z$,
obtained according to \cite{16} up to 500 GeV
(at higher energy the curves are extrapolated).

As a consequence, the flux of photons at Earth is a power law
spectrum with the same slope of the emission spectrum
up to a certain energy depending on the distance, followed by a gradual
steepening, with a sharp end at the energy $E'_{max}=E_{max}/(z+1)$.
Assuming an isotropic emission, 
the number of photons per unit area and unit energy
at the top of the atmosphere is
\begin{equation}
\frac{d\Phi_{\gamma}}{dE} = \frac{K}{4 \pi r^2} [E(z+1)]^{-\alpha} (z+1) e^{-\tau(E,z)} 
\end{equation}

$r$ is the cosmological comoving coordinate \cite{17}

\begin{equation}
r=c\frac{zq_0+(q_0-1)(-1+\sqrt{2q_0z+1})}{H_0 q_0^2 (1+z)}
\end{equation}

$H_0$ is the Hubble constant and $q_0$ the deceleration parameter
(we will use $H_0$=75 km s$^{-1}$ Mpc$^{-1}$ and $q_0$=0.5,
i.e. flat universe with $\Omega=1$).

In this simple parametrizazion L, $z$, $\alpha$ and $E_{max}$ determine
the energy fluence of gamma-rays above 1 GeV at Earth,
$F=\int_{1~GeV}^{E'_{max}}\frac{d\Phi_{\gamma}}{dE} E dE$.
The possible range of variability of these parameters can be reasonably
deduced by lower energy measurements:

\noindent
\underline{$z$}:
up to now, the distance of GRBs host galaxies has been measured 
for 6 events and the observed redshifts are respectively 
0.8, 1.0, 1.1, 1.6, 1.6, 3.4 \cite{18,19,20,21,22,23}
(a further one, GRB980425, probably associated with a Supernova 
explosion at $z$=0.0085 seems a peculiar and non standard event \cite{24});
basing on these data, we will take $z$ in the interval 0-4;

\noindent
\underline{$\alpha$}: 
most of the events observed by BATSE show power law spectra 
at energies $E>$100~KeV-1~MeV, with exponent 1.5$<\alpha<$2.5 \cite{25};
since the few GRBs detected by EGRET at energies $E>$30 MeV
also show a similar behaviour \cite{5},
one can expect a higher energy component with the same spectral
slope observed in the MeV-GeV energy region;

\noindent
\underline{$L$}: 
assuming an isotropic emission, the energies released by these 5 bursts
range between $\sim$5~10$^{51}$ and $\sim$2~10$^{54}$ ergs
in the BATSE energy range $E>$20~KeV;
it is not unreasonable to assume the energy $L$ emitted above 
1~GeV being of the same
order of magnitude (recall that a spectrum with a exponent $\alpha$=2
means equal amount of energy release per decade of energy);

\noindent
\underline{$E_{max}$}:
the value of the maximum energy of gamma-rays 
depends on unknown physical conditions and in fact is one of the
quantity that we hope to measure with this technique;
we will consider $E_{max}$ as a parameter ranging in the interval 
30~GeV-1~TeV.

\subsection{The GRB signal}

In this section we evaluate the response of a detector to a GRB,
as a function of the burst parameters and the detector characteristics.
A detector can be simply defined by the area
$A_d$ and the altitude $h$ above the sea level.
Note that the detector sensitivity does not depend on its
geometrical features, as the area of the single counters or their
relative positions, but only on the total sensitive area
(e.g. for an air shower array $A_d$ is the sum of the single counters areas).
The latitude of the detector geographic location will not be
considered due to its small effect on the background rate.

Given a GRB, defined by the parameters L, $z$, $\Delta t_e$,
$\alpha$ and $E_{max}$ previously discussed, 
and with an arrival direction corresponding to a zenith angle $\theta$,
the flux of secondary particles (number of particles per unit 
area) reaching the altitude $h$ above the sea level
induced by the gamma-rays of energy $E>$1~GeV is:

\begin{equation}
\Phi_{e^{\pm}} \ = \int_{1 \ GeV}^{E'_{max}} 
\frac{d\Phi_{\gamma}}{dE} n_e(E,\theta,h) \ dE
\end{equation}

where $n_e(E,\theta,h)$ is the mean number of particles reaching the
altitude $h$ generated by a gamma-ray of energy $E$ and 
zenith angle $\theta$. 

As an example Fig.5 shows the particles flux as a function of the altitude,
produced by a vertical GRB with $L=10^{53}$ erg, for different distances.
The spectrum is assumed with a slope $\alpha$=2 and $E_{max}$=30 GeV and 1
TeV. The flux strongly increases with the altitude; at 5000 m it
is about 3 orders of magnitude higher than at sea level.
For distances less than $z \sim$1, the flux increases with $E_{max}$,
while at higher distances, due to the absorption of the most energetic 
gamma-rays, the flux is almost independent on $E_{max}$.
The ground level flux decreases at higher zenith angles,
in particular at lower altitude (see Fig.1); 
as an example at $h$=4000 m the flux at $\theta$=30$^{\circ}$
is $\sim$3 times lower than the vertical flux, while at $\theta$=50$^{\circ}$
the flux is reduced by almost 2 orders of magnitude.

The number of particles giving a signal in the detector
\footnote{
This is true assuming that all the particles hitting a detector
give a signal; actually in a standard air shower array,
two or more particles of the same shower hitting the same counter
give only one signal, due to the limited time resolution;
however, working in the $<$ 1 TeV  energy range $n_e$ is so small 
that the fraction
of showers giving two or more particle in the same counter is negligible.}
is $N_s \ = A_d $cos$\theta \Phi_{e^{\pm}}$.
The signal must be compared to the noise 
\mbox{$\sigma _b=\sqrt{A_d  \ B(h) \  \Delta t}$},
given by the background fluctuations during the time 
$\Delta t=\Delta t_e (z+1)$,
where $B(h)$ is background rate (number of events per unit area and unit time).
A GRB is detectable with a statistical significance
of $n$ standard deviations if $N_s/\sigma _b > n$.
The minimum value of $n$ necessary to consider an excess as a significant
signal depends on the search strategy. In the case of a search
in coincidence with satellites, since the frequency of GRBs occurring
in the field of view of a ground based detector is about one event every
few days, a level of 4 standard deviations is sufficient to give a high 
reliability to the observation. In the following discussion we set $n=$ 4. 
A detector of a given area $A_d$ and altitude $h$ is therefore sensitive to 
GRBs with a ground level particle flux 

\begin {equation}
\Phi_{e^{\pm}} > \frac{n}{cos \theta} \sqrt{\frac{B(h) \Delta t}{A_d}}
\end {equation}

As an example, in Fig.5 the minimum detectable ground level flux 
as a function of the altitude
is shown for a detector of area $A_d$=10$^3$ m$^2$ and a GRB with a
time duration $\Delta t$=1 s and zenith angle $\theta$ = 0$^{\circ}$.
The minimum flux ranges from $\sim$2 to $\sim$6 particles m$^2$ for altitudes
from the sea level to 5000 m.

Conversely, if one aims to be sensitive to a certain ground level
flux $\Phi_{e^{\pm}}$, the requirement for the minumum detector area
is  

\begin {equation}
A_d > \frac{n^2 B(h) \Delta t}{\Phi_{e^{\pm}}^2 cos^2 \theta}
\end {equation}

To give an estimate of the magnitude of the largest 
fluxes that one can reasonably expect,
we have extrapolated the power law fits of 15 GRBs spectra
obtained by the TASC instrument of EGRET \cite{5} up to a maximum energy
$E_{cut}$,
and we have calculated the secondary particle flux at the ground level
using equation (3) (setting $E'_{max}=E_{cut}$ and 
$d\Phi_{\gamma}/dE$ equal to the EGRET spectra).
The maximum flux is given by GRB910709 ($\alpha$=1.76), which would
produce a flux $\Phi_{e^{\pm}} \sim$450(6500) particles m$^{-2}$ at 
$h$=2000(5000) m
assuming $E_{cut}$=1 TeV, and $\Phi_{e^{\pm}} \sim$12(420) particles m$^{-2}$
for $E_{cut}$=30 GeV.
In total, the number of GRBs (out of 15) giving
a flux $\Phi_{e^{\pm}}>$ 10 particle m$^{-2}$ at the altitude $h$=2000(5000) m
is $N=$ 3(10) for $E_{cut}$=1 TeV, and $N=$1(4) for $E_{cut}$=30 GeV. 
As expected the ground level signals are very sensitive to the maximum energy
$E_{cut}$, that is determined by the maximum emitted energy at the source
and by the absorption in the intergalactic space.
Our analysis of the EGRET events indicates that even if the GRBs spectra
at Earth do not extend to very high energy, the most powerful events
can give ground level fluxes at mountain altitudes 
$\Phi_{e^{\pm}} >$ 10 particles m$^{-2}$.
The minimum sensitive area required to detect a flux of 10 particles m$^{-2}$
can be deduced by the expression (5)
using the background estimates given in Fig.3.
For example, at $h$=4000 m, 
\mbox{$A_{min}=214 \frac{\Delta t}{1 s} \frac{1}{cos^2 \theta}$ m$^2$.}

So far we have expressed the sensitivity of the single particle
technique in terms of the ground level particles flux.
To discuss the sensitivity 
in terms of the primary gamma-rays flux at the top of the atmosphere, 
it is useful to introduce $F_{min}$,
the minimum energy fluence of gamma-rays above 1 GeV
required to give a detectable signal.
We considered a detector of area $A_d$=10$^3$ m$^2$
and a GRB with a zenith angle $\theta$ = 0$^{\circ}$,
$\Delta t$=1 s, and $z$ sufficiently small to make the distortions of the
spectrum (absorption and redshift) negligible.
Fig.6 shows the minimum energy fluence $F_{min}$
in the range 1 GeV-$E_{max}$ 
as a function of $h$, for different spectral parameters.
If the detector is located at very high mountain altitude (h$>$ 4000 m)
fluences of few 10$^{-5}$ erg cm$^{-2}$ are observable 
also from soft spectra ($\alpha$=2.5, $E_{max}$=30 GeV),
while fluences of few 10$^{-6}$ erg cm$^{-2}$ are detectable only
from hard spectra ($\alpha$=1.5, $E_{max}$=1 TeV).
The minimum observable fluence for a detector with a generic
area $A_d$ = $k$ 10$^3$ m$^2$ and for a GRB with a generic time duration
$\Delta t=T$ s, scales as $\sqrt{T/k}$.

It is worth noting that in the evaluation of the sensitivity
we have not taken into account the possibility
to increase the GRB signal by converting a fraction of photons in detectable
charged particles as discussed in Section 2.

\subsection{Expected rate of observations}

A rough evaluation of the expected rate of positive observations,
as a function of the unknown spectral parameters at high energies,
can be performed by making assumptions on the luminosity function and 
the spatial distribution of GRBs in the universe.

As a first step we evaluate the probability to detect a GRB
with a given luminosity\footnote{
The luminosity is usually defined as the amount of energy emitted
per unit time, while here we call luminosity 
the energy emitted during the total emission time $\Delta t_e$.}
L as a function of $z$.
Due to the absorption, the luminosity $L$ required to observe a GRB 
increases with the
distance more rapidly than what is expected by simple geometrical effects.
Fig.7 shows the minimum luminosity $L$ necessary to make a burst
detectable 
by a 10$^3$ m$^2$ detector located at different altitudes,
as a function of the redshift $z$.
The burst is assumed at a zenith angle $\theta$=0$^{\circ}$, 
with a time duration
$\Delta t_e$=1 s, $\alpha$=2 and $E_{max}$=30 and 1000 GeV.
Obviously, for a detector of a generic area $A_d$ = $k$ 10$^{3}$ m$^2$ 
the minimum observable luminosity scales as 1/$\sqrt{k}$.
From these curves  one can deduce the maximum distance
observable by the detector, for a given $L$.
As an example, if $L$ = $10^{51}$(10$^{54}$) erg and $E_{max}$=1 TeV,
a 10$^3$ m$^2$ detector at the
sea level could see the burst if the source is 
located within a distance $z \sim$ 0.003(0.4)
while a detector at $h$=4 km could see up to $z \sim$ 0.2 (3.0).

As a matter of fact not all GRBs with a given $L$ are
observable if $z$ is less than a certain value: the detection depends 
on the zenith angle $\theta$. A correct approach to the problem 
is to calculate the probability $P(L,z)$ to observe a burst of
luminosity $L$ and distance $z$ taking into account all possible
arrival directions

\begin{equation}
P(L,z) = \frac{1}{4 \pi} \int_{0}^{2\pi} d\phi \int_{0}^{\pi} J(L,z,\theta) sin\theta d\theta
\end{equation}

where $J(L,z,\theta)$=1 if the burst is detectable, i.e. if the 
statistical significance
of the signal is larger then the required value, otherwise
$J(L,z,\theta)$=0.

As a second step we evaluate
the fraction $\epsilon(L)$ of observable events over the total number of 
GRBs of luminosity $L$ distributed in the universe. 
For this purpose one should know
the form of the GRBs density $n(z,L)$ (number of bursts 
per unit volume and unit time). We will adopt the simplest model:
$a)$ GRBs are distributed in the universe with constant 
density and frequency
in a comoving coordinate frame up to a maximum distance $z_{max}$,
$b)$ GRBs show no evolution, i.e. the luminosity distribution is
independent on $z$. 

With these assumptions
$n(z,L)\propto (1+z)^3/(1+z)$; 
the term $(1+z)^3$ takes into account that the universe
at a time corresponding to a redshift $z$ was smaller than now by a factor 
$(1+z)$; the term $1/(z+1)$ represents the frequency decrease due 
to time dilation. The fraction of observable events up within the distance
$z_{max}$ is

\begin{equation}
\epsilon(L) = \frac{\int_{0}^{z_{max}} P(L,z) (1+z)^2 \frac{dV}{dz} dz}
{\int_{0}^{z_{max}} (1+z)^2 \frac{dV}{dz} dz}
\end{equation}

where 
$dV(z)$ is the volume element, according to \cite{17}

\begin{equation}
dV(z) = \frac{4 \pi c^3}{(1+z)^6 H_0^3 q_0^4} (1+2q_0z)^{-1/2} 
[zq_0+(q_0-1)(-1+\sqrt{2q_0z+1})]^2 dz
\end{equation}

Fig.8 shows $\epsilon(L)$ for a 10$^3$ m$^2$ detector at different
altitudes, calculated setting $z_{max}$=4. 
The burst is assumed with an emission time
$\Delta t_e$=1 s, a slope $\alpha$=2 and $E_{max}$=30, 100 and 1000 TeV. 
The corresponding fraction of observable events for a detector area
$A_d$ = $k$ 10$^{3}$ m$^2$ is $\epsilon(L \sqrt{k})$.
It is interesting to note that, for a given detector altitude,
a GRB with $E_{max}$=100 GeV becomes more easily observable than
a GRB with the same luminosity and $E_{max}$=1 TeV,
if $L$ is larger than a certain value; this is due to the fact 
that increasing the luminosity one can observe at
larger distances and consequently the absorption of high energy photons
becomes more important.

The final step is the calculation of the total rate 
of observable events, considering all the possible luminosities $L$

\begin{equation}
R_{obs} = R_{tot}\int_{L_{min}}^{L_{max}} \frac{dN(L)}{dL} \epsilon(L) dL 
\end{equation}

where $R_{tot}$ is the total bursts rate in the universe within a
distance $z_{max}$ and dN(L)/dL is the luminosity distribution of GRBs.
A rough evaluation of $R_{tot}$ is given from BATSE data.
Since BATSE observes $\sim$1 burst per day, with a detection 
efficiency of  $\sim$0.3, the rate deduced from BATSE
obervations is $R_B\sim$1000 y$^{-1}$. In the assumption that BATSE
can observe GRBs up to a distance $z \sim$ 4, we set $R_{tot}=R_B$.

The luminosity distribution is a more troublesome topic:
in the KeV-MeV energy region
a large number of studies has been done in order to deduce a luminosity
function suitable to reproduce the flux distribution of BATSE 
bursts but a general agreement about the subject  doesn't exist.
Different forms of $N(L)$ could reproduce the BATSE data and 
so far there exist too few luminosity measurements to put
stringent constraints on the distribution.
One point is clear: the 5 events whose distance has been measured
show that GRBs are not standard candles and the luminosity distribution
at low energy is a very broad function, ranging over almost 4 orders of 
magnitude.
Lacking more compelling data we will assume the luminosity
function in the form $dN(L)/dL=C L^{-\beta}$
with $L_1 < L < L_2$; tentatively,
we set $L_1 = 10^{51}$ erg and $L_2 = 10^{55}$ erg,
and consider $\beta$ as parameter ranging in the interval 1-4.

Fig.9 shows the expected rate of observation as a function of $\beta$
for a 10$^3$ m$^2$ detector at different altitudes.
The burst spectrum is assumed with $\alpha$=2 and  
$E_{max}$=30, 100 and 1000 GeV.
The rates are given for the emission times $\Delta t_e$=1 s and $\Delta t_e$=10 s.
The curves show that a 10$^3$ m$^2$ detector located at 
$h \sim$ 4 km, could observe a number of bursts per year ranging from $\sim$ 1 to 
20-30 if the slope of the luminosity function is not steeper than
2-2.5.

\section{Conclusions}

The single particle technique provides a simple method of observing
GRBs in the energy range 1-10$^3$ GeV using large air shower
arrays located at high mountain altitude.
The observation of a significant excess in the counting rate in coincidence
with a burst seen by a satellite
would provide, in a very simple and economical way,
important informations on the GRB high energy component.
Besides the study of the temporal behaviour of the high energy emission,
the single particle observation could allow the evaluation of
the total energy $L$ emitted above 1 GeV (or equivalently
the maximum energy of the spectrum $E_{max}$) provided that
the spectral slope is deduced by a complementary 
lower energy observation and the redshift is measured with optical instruments.

The expected rate of observation depends on bursts parameters that are
so far poorly known, as the high energy spectrum shape,
the distribution of the sources in the universe, and in
particular the luminosity function. 
However, making reasonable assumptions,
we conclude that a detector of  $A_d \sim 10^3$ m$^2$ at
an altitude of $\sim$ 4000 m above the sea level 
could be expected to detect 
at least few events per year if 
the GRB spectrum extends up to $\sim$ 30 GeV
and the luminosity function slope is not too large. 

An attempt to detect high energy GRBs using the
single particle technique was performed 
by G.Navarra about 15 years ago using a 
small detector located at Plateau Rosa at 3500 m a.s.l.\cite{26}.
At present two experiments are using this technique:
the 350 m$^2$ air shower array EASTOP,
working at 2000 m a.s.l. at the Gran Sasso Laboratory (Italy)
and INCA, a 48 m$^2$ detector
located at Mount Chacaltaya (Bolivia) at 5200 m a.s.l.

EASTOP has been searching high energy 
GRBs since 1991, showing the possibility 
to perform the single particle measurement with the
necessary stability over long periods of time, monitoring
with very good reliability short time fluctuations in the counting rate.
In a search for high energy gamma-rays in coincidence
with about 300 BATSE events detected during 1992-1998, EASTOP
obtained upper limits to the 10 GeV-1 TeV fluence
as low as $F \sim 10^{-4} - 5 \ 10^{-3}$ erg cm$^{-2}$, 
for small zenith angle GRBs ($\theta<30^{\circ}$)\cite{27}.
In a similar search performed with about 2 years of data,
INCA obtained upper limits to the 1 GeV-1 TeV fluence
$F \sim$ 5 $10^{-5}-10^{-4}$ erg cm$^{-2}$,
also for small zenith angle GRBs\cite{28,29}. The limited area of INCA is
highly compensated by its extreme altitude.

The single particle technique has been adopted as well by the CYGNUS 
collaboration using water Cerenkov detectors, with a total sensitive
area of $\sim$210 m$^2$ at an altitude of 2130 m. They reported
upper limits on the 1~GeV-30~TeV energy fluence $F\sim
10^{-4} - 2 \ 10^{-1}$ erg cm$^{-2}$ s$^{-1}$,
in coincidence with 9 strong GRBs with zenith angles up to 60$^{\circ}$ 
\cite{30}.

Good possibilities of positive detections
come from the new generation air shower arrays
consisting of very large sensitive surfaces, as ARGO and MILAGRO,
conceived to detect small air showers with the aim of observing 
gamma-ray sources at energy E $<$ 1 TeV. 
ARGO, under construction at the
Yangbajing Laboratory (Tibet) at 4300 m, will consist of a carpet of 7500 m$^2$
of Resistive Plate Chambers, covered by a layer of lead 0.5 cm thick to
convert a fraction of shower photons and 
increase the number of detectable particles \cite{31}.
MILAGRO, a Cerenkov detector using a 4800 m$^2$ pond of water,
is located at 2600 m, near Los Alamos. An interesting feature
of the water Cerenkov technique is the possiblity to perform a calorimetric
measurement of the shower particles and to detect most of
the photons, due to their interactions in the water \cite{30,32}.
The two future experiments have about the same sensitivity in observing
GRBs by using the single particle technique. The lower altitude
of MILAGRO, that reduces its sensitivity by a factor $\sim$3 with
respect to ARGO,
is compensated by the higher capability of detecting the shower photons.
If the assumptions that we have made on high energy GRBs are correct, 
both the experiments likely will detect at least few events per year.

\blankline

{\bf\large Acknowledgements}

\blankline

We are grateful to P.Lipari, G.Navarra, O.Saavedra and P.Vallania 
for helpful discussions and comments.


\newpage



\newpage

{\bf\large Figure captions}

\blankline

{\bf Fig.1} Mean number of charged particles reaching different
altitudes $h$ above the sea level (in km), generated in the atmosphere 
by a photon of zenith angle
$\theta$ = 0$^{\circ}$ (full line) and $\theta$ = 30$^{\circ}$ (dashed line), 
as a function of the photon energy.

\blankline

{\bf Fig.2} Ratio of the number of photons
and the number of $e^{\pm}$ reaching the ground, as a function of the
altitude. The points are given for different gamma-ray energies:
10 GeV (circles), 100 GeV (squares), 1 TeV (triangles).
The photon energy threshold is 3 MeV.

\blankline

{\bf Fig.3} Background rate due to secondary charged particles as a function
of the altitude. The kinetic energy threshold used in the simulation
is 3 MeV for electrons and photons and 50 MeV for muons 
and hadrons.

\blankline

{\bf Fig.4}
A power law gamma-ray spectrum with a slope $\alpha$=2 affected by 
$\gamma + \gamma \rightarrow e^+ e^-$ pair production,
for different redshifts of the source.

\blankline

{\bf Fig.5}
Ground level secondary particles flux 
produced by a vertical GRB with $L=10^{53}$ erg, located at different distances
$z$, as a function of the altitude.
The GRB spectrum is assumed with a slope $\alpha$=2 and 
$E_{max}$=1 TeV (solid line) and $E_{max}$=30 GeV (dotted line).
As an example of a detector sensitivity,
the dashed line represents the minimum ground level flux observable
by a detector of area $A_d$=1000 m, from a GRB with $\Delta t$=1 s
and zenith angle $\theta$ = 0$^{\circ}$.
\blankline

{\bf Fig.6}
Minimum energy fluence in the range 1~GeV - $E_{max}$ detectable
by a 10$^3$ m$^2$ detector as a function of the altitude. 
The curves are given for different GRB spectral slopes $\alpha$ and 
$E_{max}$ values: curves $a$ correspond to $\alpha$=2.5, curves $b$ to
$\alpha$=1.5; subscripts 1,2,3 refers
respectively to $E_{max}$=30, 100, 1000 GeV.

\blankline

{\bf Fig.7}
Minimum amount of energy $L$ released above 1 GeV by a GRB
at distance $z$ necessary to make the event observable by a 
10$^3$ m$^2$ detector located at different altitudes $h$ (in km), 
as a function of $z$. The curves are
calculated for a GRB spectrum with slope $\alpha$=2 and 
$E_{max}$=1 TeV (solid line) and $E_{max}$=30 GeV (dashed line).

\blankline

{\bf Fig.8}
Fraction of GRBs observable up to a distance $z_{max}$=4
as a function of the luminosity $L$, for a 
10$^3$ m$^2$ detector located at different altitudes $h$ (in km).
The curves are
calculated for a GRB spectrum with slope $\alpha$=2 and 
$E_{max}$= 1 TeV (solid line), 100 GeV (dashed line)
and 30 GeV (dotted line).

\blankline

{\bf Fig.9}
Expected rate of observable GRBs as a function of the slope
$\beta$ of the luminosity function,
for a 10$^3$ m$^2$ detector located at different altitudes $h$ (in km).
The curves are
calculated for a GRB spectrum with a slope $\alpha$=2 and 
$E_{max}$=1 TeV (solid line), 100 GeV (dashed line)
and 30 GeV (dotted line). The emission time is $\Delta t_e$ = 1 s (upper figure)
and 10 s (lower figure).

\newpage

\begin{figure}[ht]
 \begin{center}
\mbox{\psfig{file=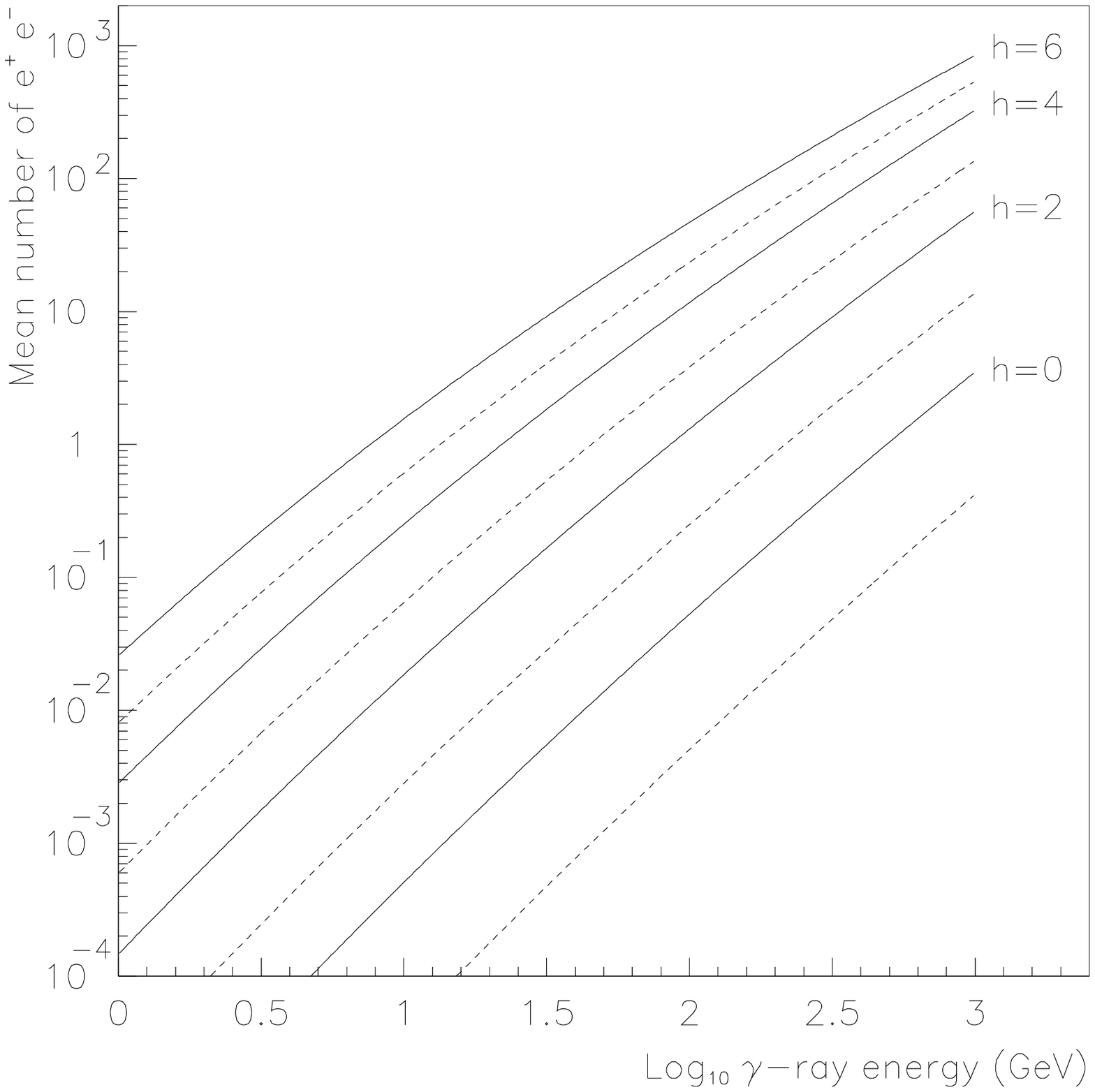,height=15.cm,width=15.cm,bbllx=0bp,bblly=0bp,bburx=525bp,bbury=600bp,clip=}}
 \end{center} 
\vspace{2cm}
\caption{ }
\end{figure}

\newpage

\begin{figure}[ht]
 \begin{center}
\mbox{\psfig{file=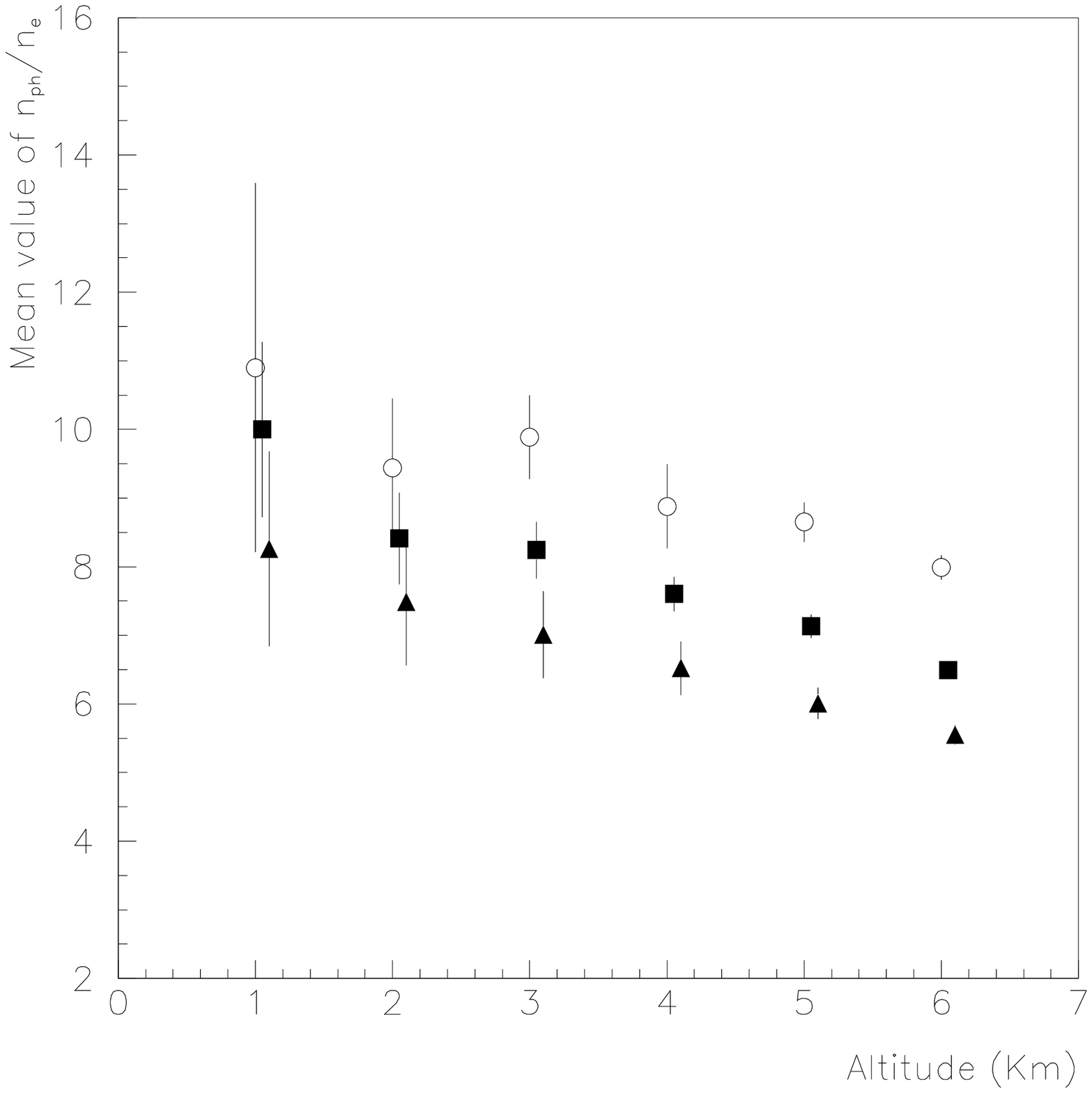,height=15.cm,width=15.cm,bbllx=0bp,bblly=0bp,bburx=525bp,bbury=600bp,clip=}}
 \end{center} 
\vspace{2cm}
\caption{ }
\end{figure}

\newpage

\begin{figure}[ht]
 \begin{center}
\mbox{\psfig{file=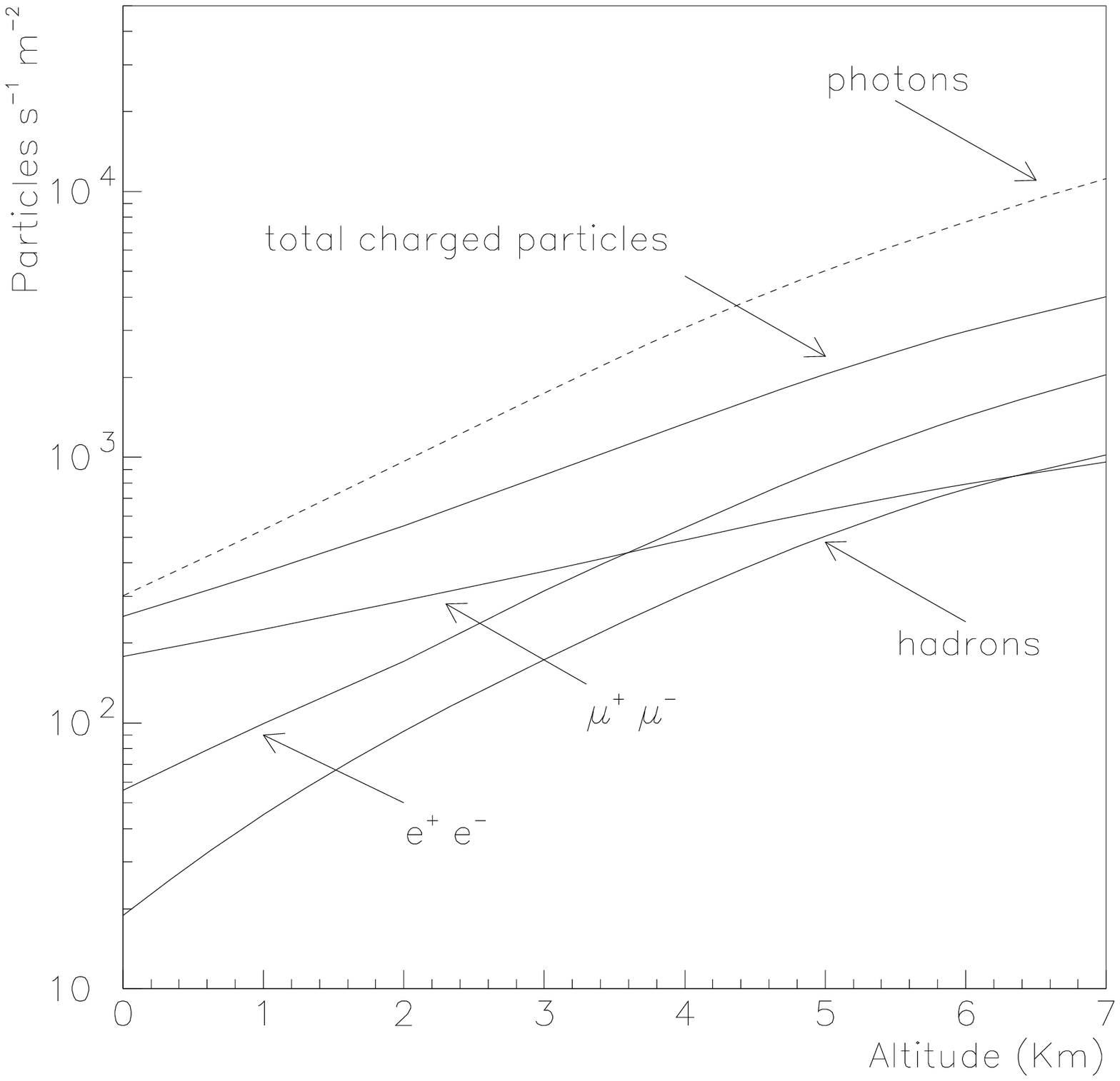,height=15.cm,width=15.cm,bbllx=0bp,bblly=0bp,bburx=525bp,bbury=600bp,clip=}}
 \end{center} 
\vspace{2cm}
\caption{ }
\end{figure}

\newpage

\begin{figure}[ht]
 \begin{center}
\mbox{\psfig{file=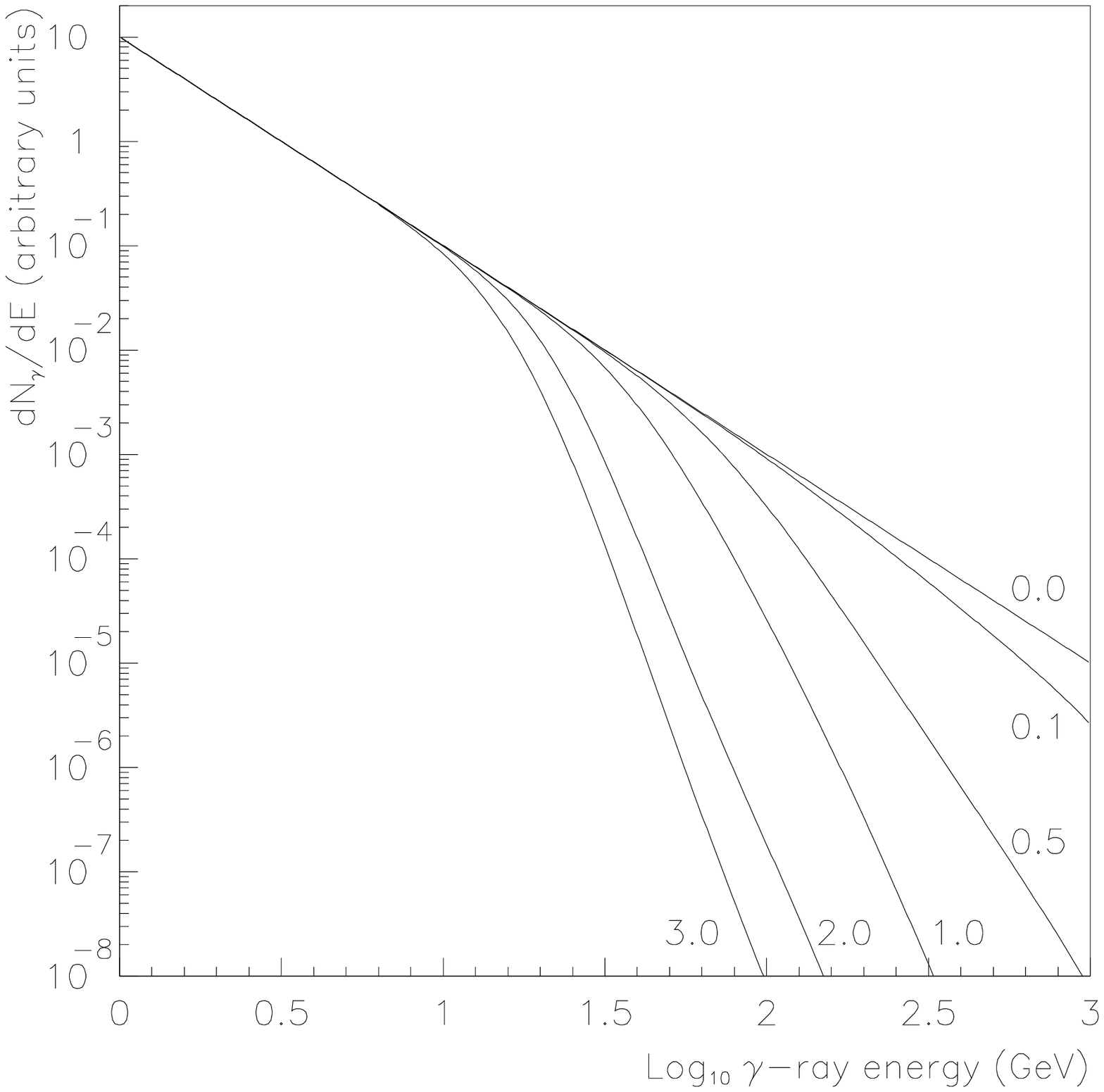,height=15.cm,width=15.cm,bbllx=0bp,bblly=0bp,bburx=525bp,bbury=600bp,clip=}}
 \end{center} 
\vspace{2cm}
\caption{ }
\end{figure}

\newpage

\begin{figure}[ht]
 \begin{center}
\mbox{\psfig{file=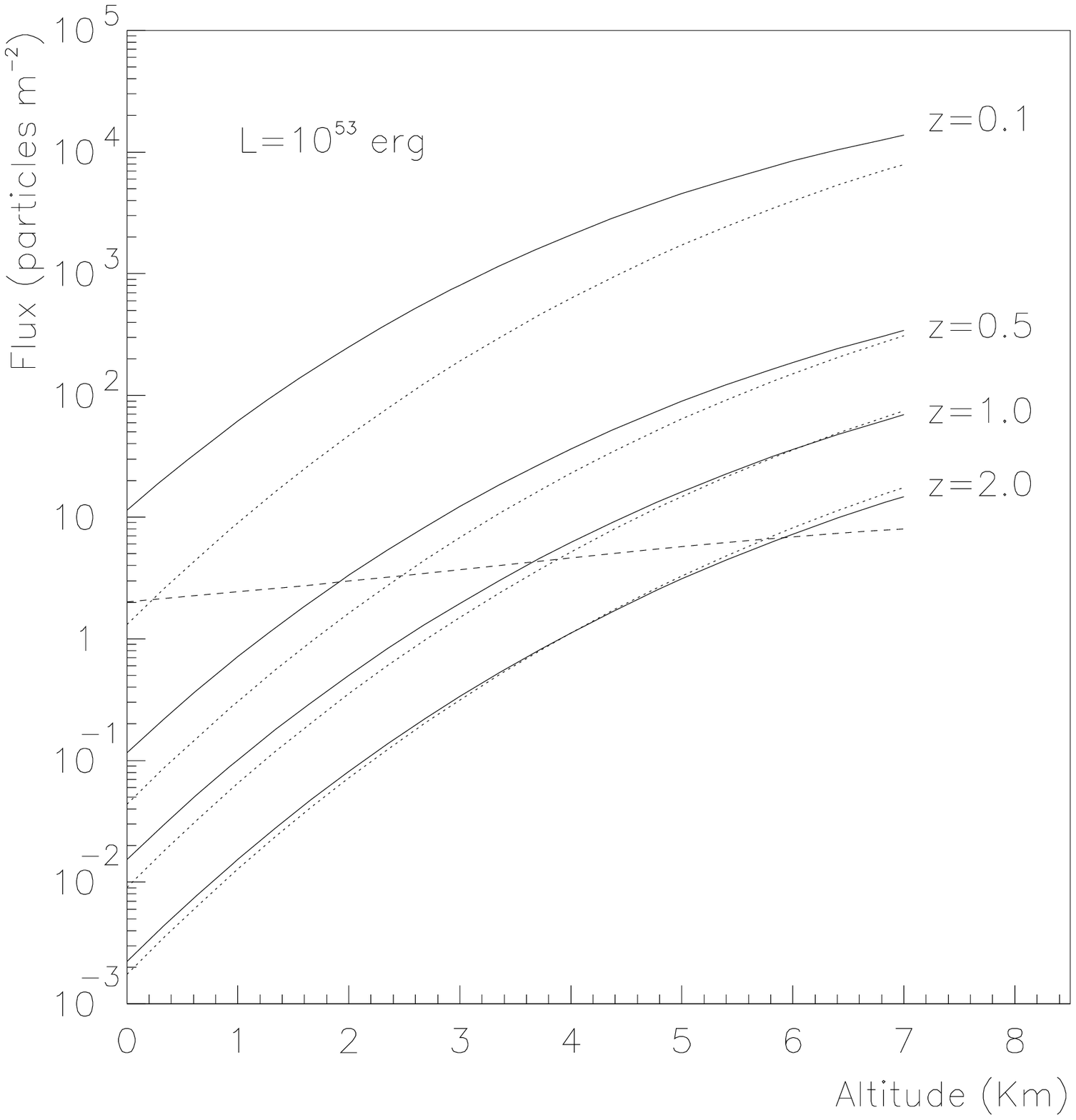,height=15.cm,width=15.cm,bbllx=0bp,bblly=0bp,bburx=525bp,bbury=600bp,clip=}}
 \end{center} 
\vspace{2cm}
\caption{ }
\end{figure}

\newpage

\begin{figure}[ht]
 \begin{center}
\mbox{\psfig{file=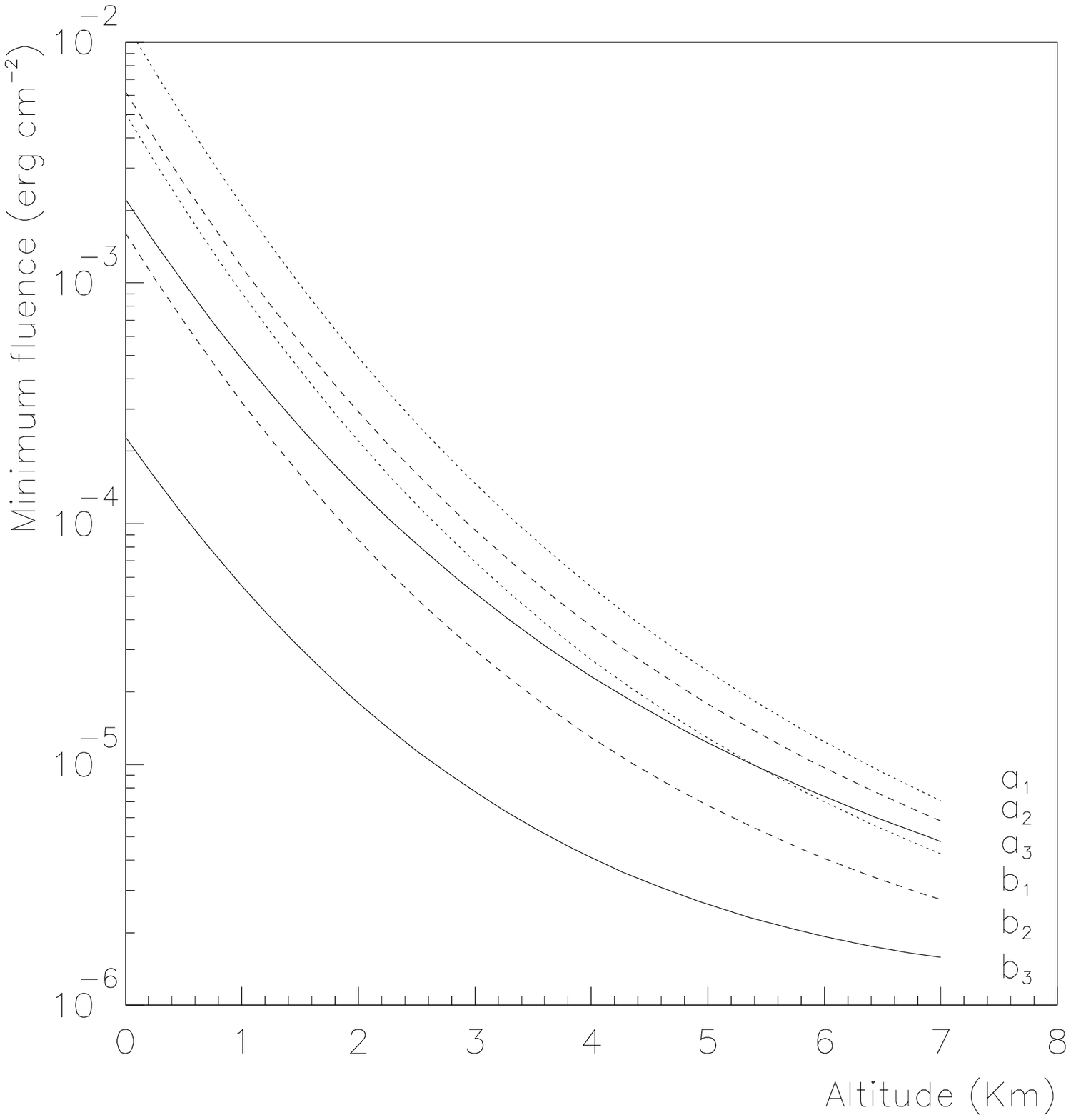,height=15.cm,width=15.cm,bbllx=0bp,bblly=0bp,bburx=525bp,bbury=600bp,clip=}}
 \end{center} 
\vspace{2cm}
\caption{ }
\end{figure}

\newpage

\begin{figure}[ht]
 \begin{center}
\mbox{\psfig{file=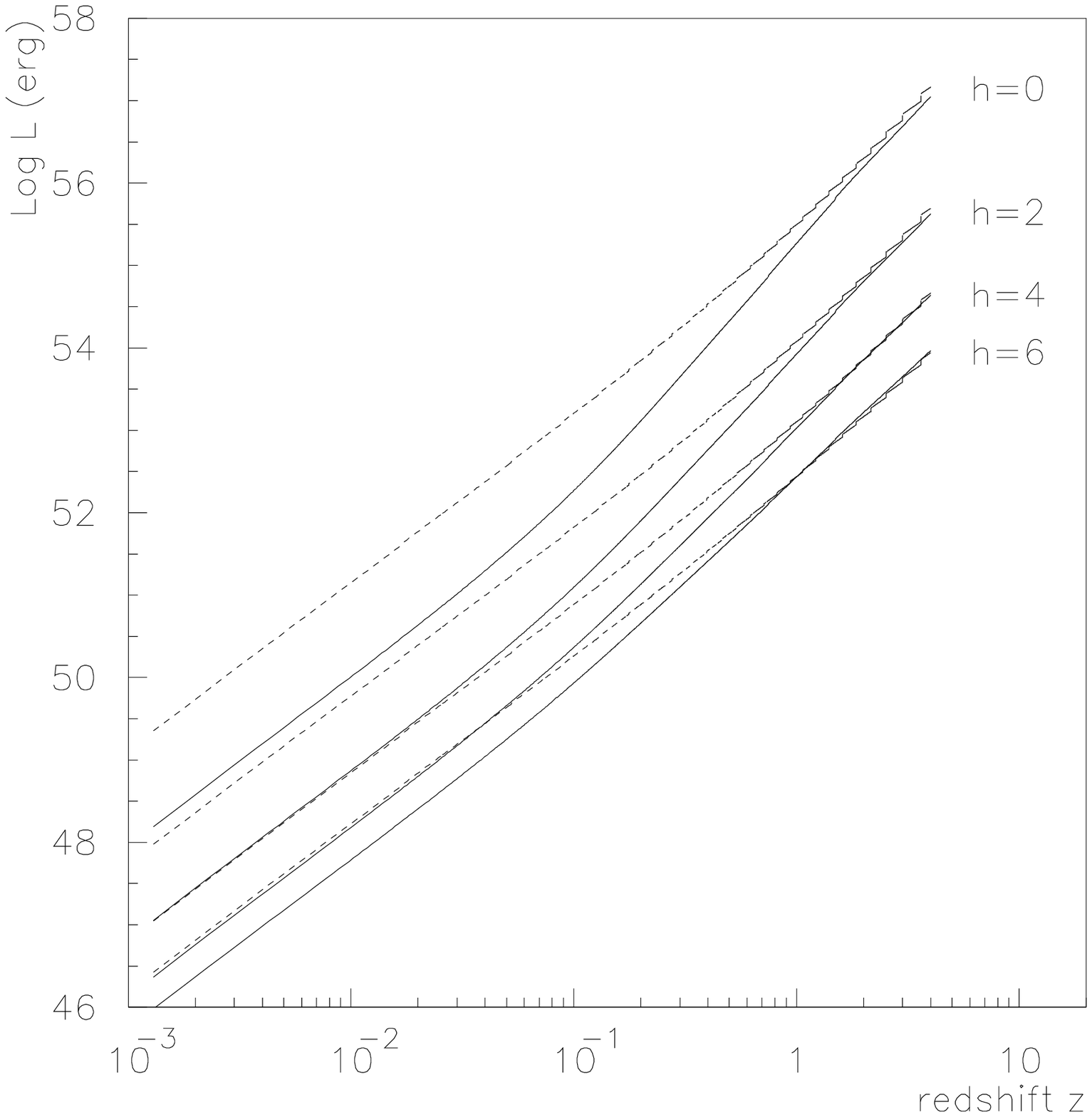,height=15.cm,width=15.cm,bbllx=0bp,bblly=0bp,bburx=525bp,bbury=600bp,clip=}}
 \end{center} 
\vspace{2cm}
\caption{ }
\end{figure}

\newpage

\begin{figure}[ht]
 \begin{center}
\mbox{\psfig{file=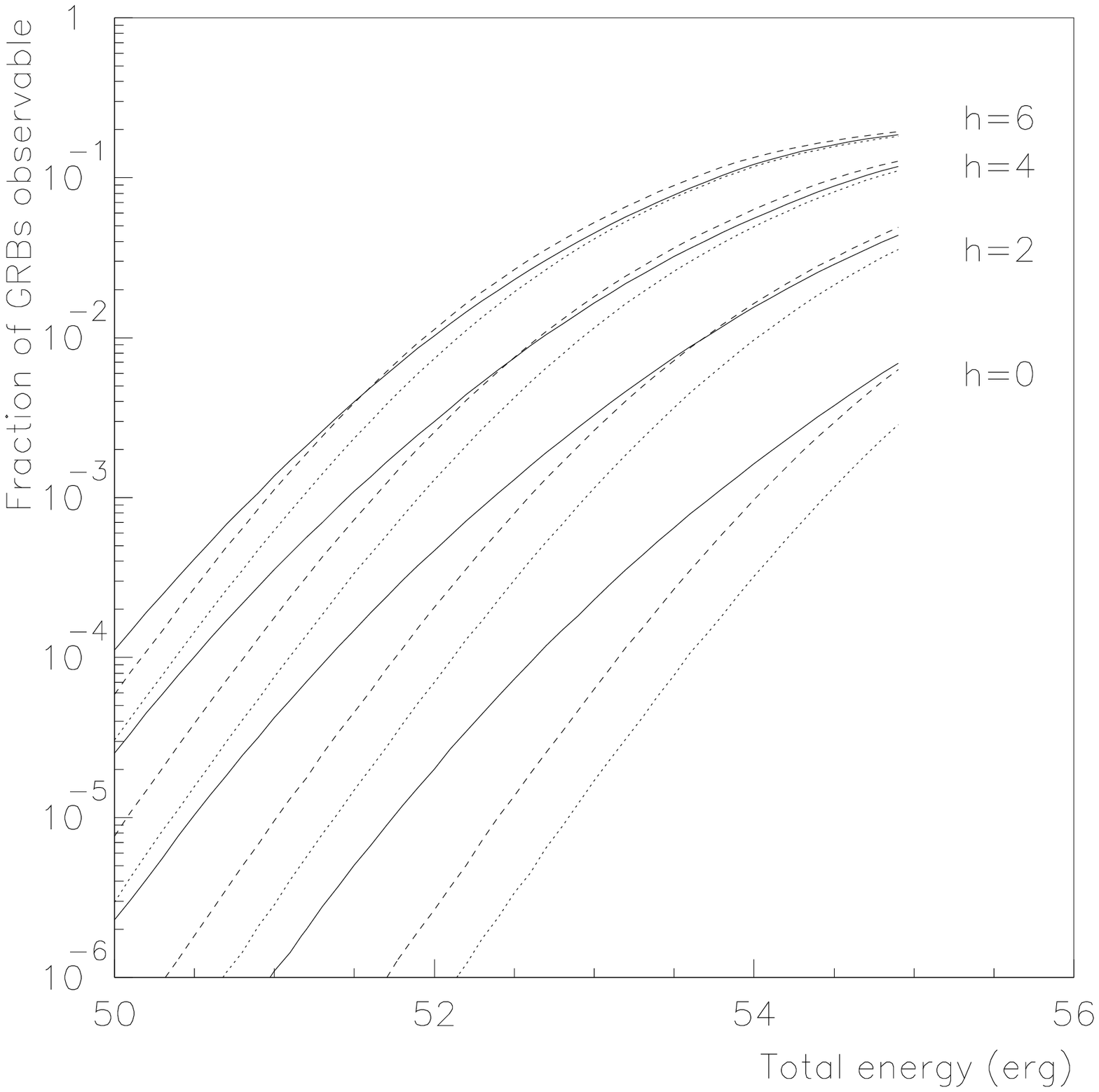,height=15.cm,width=15.cm,bbllx=0bp,bblly=0bp,bburx=525bp,bbury=600bp,clip=}}
 \end{center} 
\vspace{2cm}
\caption{ }
\end{figure}

\newpage

\begin{figure}[ht]
 \begin{center}
\mbox{\psfig{file=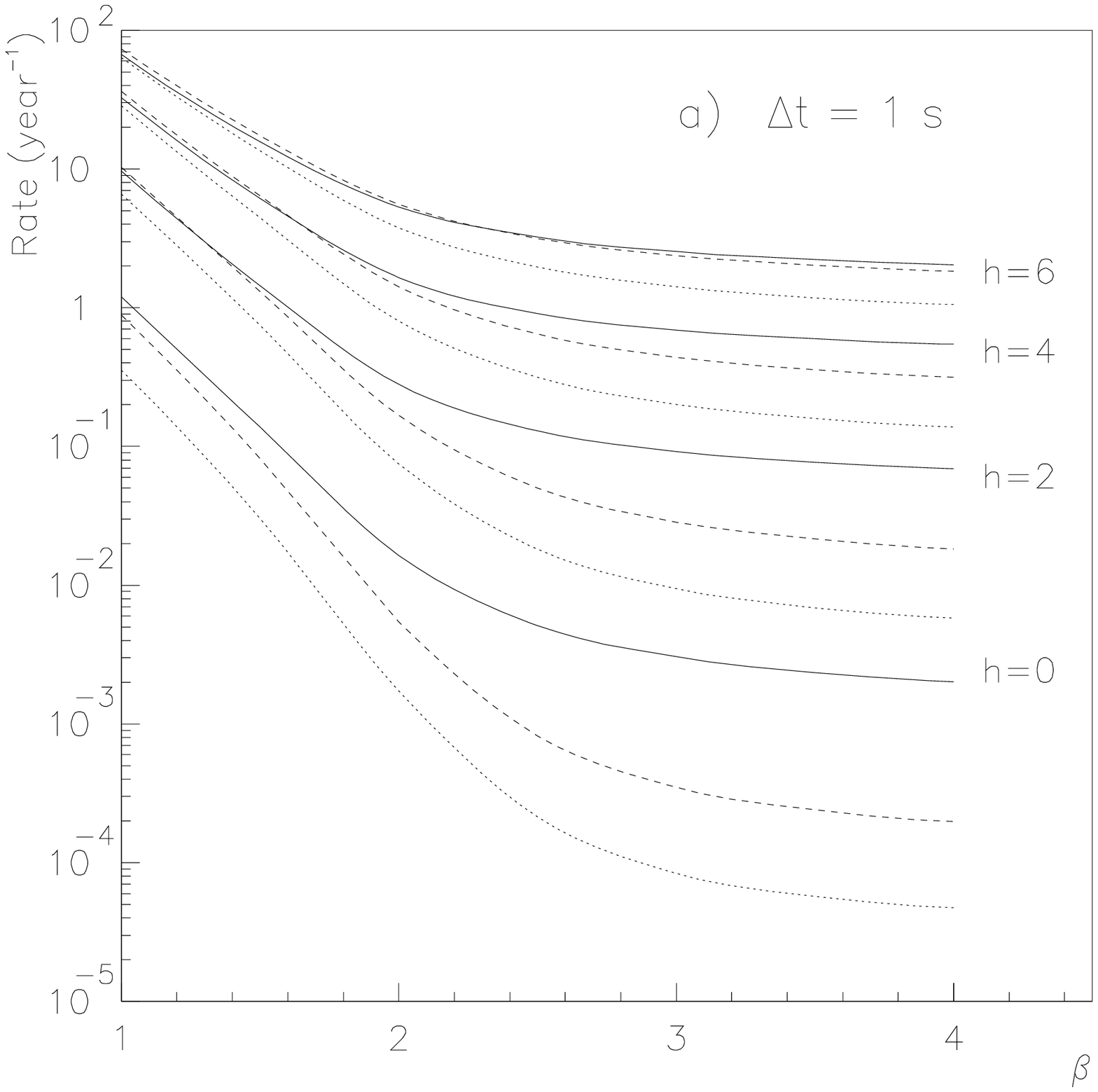,height=9.cm,width=9.cm,bbllx=0bp,bblly=0bp,bburx=525bp,bbury=600bp,clip=}}
\mbox{\psfig{file=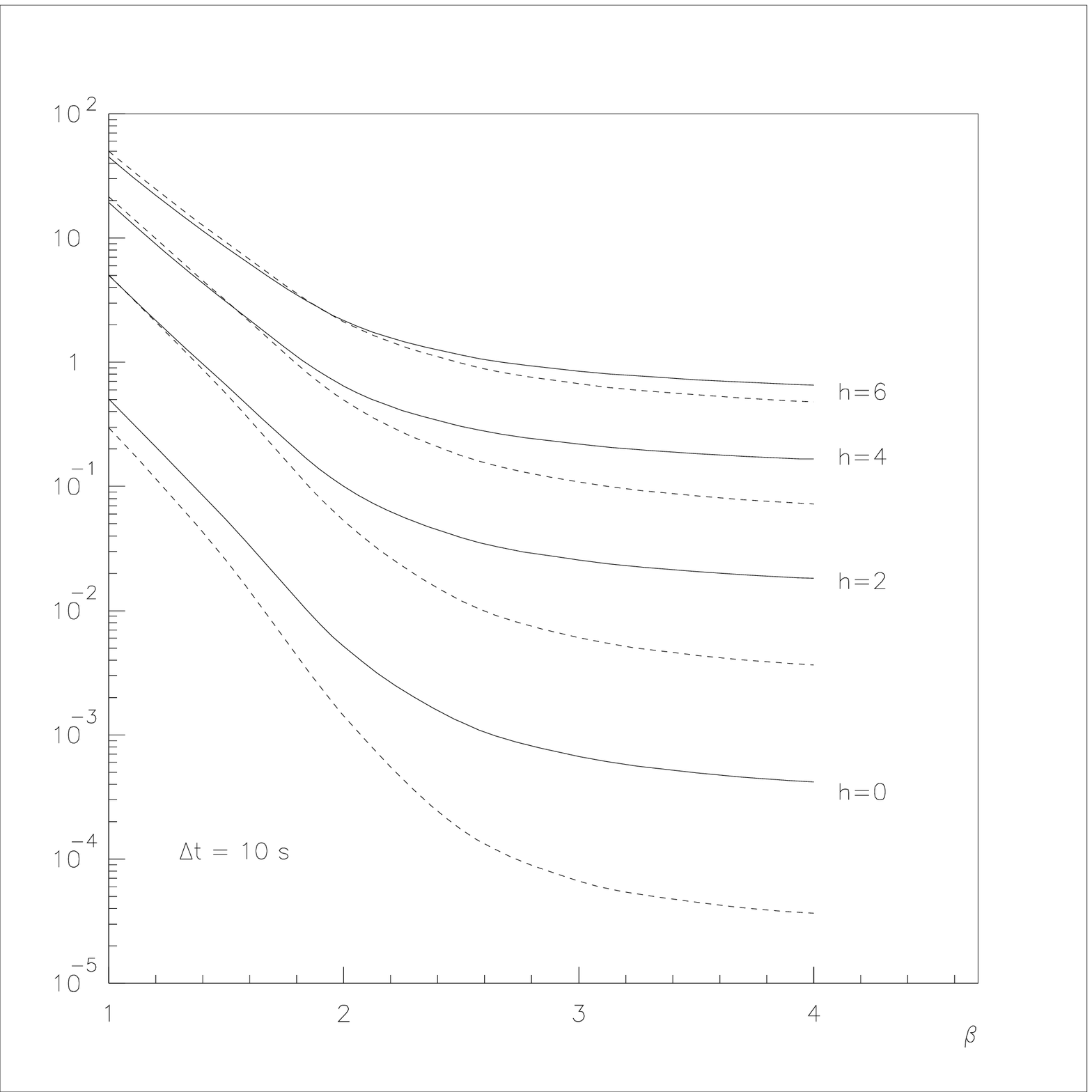,height=9.cm,width=9.cm,bbllx=0bp,bblly=0bp,bburx=525bp,bbury=600bp,clip=}}
 \end{center} 
\vspace{2cm}
\caption{ }
\end{figure}

\end{document}